\documentclass{optica-article}

\journal{opticajournal} 

\articletype{Research Article}

\usepackage{lineno}

\usepackage{hyperref}
\usepackage{float}
\usepackage{graphicx}
\usepackage{subcaption}
\usepackage{caption}
\usepackage{microtype}
\usepackage{booktabs}
\usepackage[inter-unit-product={\;},output-decimal-marker={.},per-mode=symbol-or-fraction,sticky-per,exponent-product = \cdot, output-product = \cdot,separate-uncertainty=true]{siunitx}
\usepackage[titletoc]{appendix}
\usepackage[wby]{callouts}
\usepackage{amsfonts}

\begin{document}

\title{Fabrication of low-loss III-V Bragg-reflection waveguides for parametric down-conversion}

\author{Hannah Thiel,\authormark{1,*} Marita Wagner,\authormark{1,2,3} Bianca Nardi,\authormark{1} Alexander Schlager,\authormark{1} Robert J. Chapman,\authormark{1,4} Stefan Frick,\authormark{1} Holger Suchomel,\authormark{5} Martin Kamp,\authormark{5} Sven Höfling,\authormark{5} Christian Schneider,\authormark{5,6} and Gregor Weihs\authormark{1}}

\address{\authormark{1}Institut f\"{u}r Experimentalphysik, Universit\"{a}t Innsbruck, 6020 Innsbruck, Austria\\
\authormark{2}CIC biomaGUNE, Basque Research and Technology Alliance, 20014 Donostia-San Sebastian, Spain\\
\authormark{3}CIC nanoGUNE, Basque Research and Technology Alliance, 20018 Donostia-San Sebastian, Spain\\
\authormark{4}Optical Nanomaterial Group, Institute for Quantum Electronics, Department of Physics, ETH Zurich, 8093 Zurich, Switzerland\\
\authormark{5}Technische Physik, Universit\"{a}t W\"{u}rzburg, 97074 W\"{u}rzburg, Germany\\
\authormark{6}Institute of Physics, University of Oldenburg, 26129 Oldenburg, Germany\\}

\email{\authormark{*}hannah.thiel@uibk.ac.at} 


\begin{abstract*} 
Entangled photon pairs are an important resource for quantum cryptography schemes that go beyond point-to-point communication.
Semiconductor Bragg-reflection waveguides are a promising photon-pair source due to mature fabrication, integrability, large transparency window in the telecom wavelength range, integration capabilities for electro-optical devices as well as a high second-order nonlinear coefficient.
To increase performance we improved the fabrication of Bragg-reflection waveguides by employing fixed-beam-moving-stage optical lithography, low pressure and low chlorine concentration etching, and resist reflow.
The reduction in sidewall roughness yields a low optical loss coefficient for telecom wavelength light of \( \alpha_\mathrm{reflow}=0.08\left(6\right)\,\mathrm{mm}^{-1}\).
Owing to the decreased losses, we achieved a photon pair production rate of \(8800\left(300\right)\,\left(\mathrm{mW\cdot s\cdot mm}\right)^{-1}\) which is 15-fold higher than in previous samples.

\end{abstract*}

\section{Introduction}
Nonlinear optics has a multitude of applications, such as all-optical switching, efficient detection, and multiplexing for increased data transfer rates in fiber networks.
In addition to its classical applications, nonlinear optics has strong applications in quantum communication.
Protocols that go beyond point-to-point quantum cryptography, like device-independent schemes and quantum repeaters, rely on entangled photon pairs \cite{Schwonnek2021,Li2019}.
A leading method to produce entangled photons is by parametric down-conversion.
As the need for mass-deployable devices with increased complexity grows, integrable on-chip systems become the only practical solution.
While the significance of miniaturized systems is undisputed, it remains uncertain what would constitute the ideal platform.

An ideal photon-pair source has to offer integration capabilities in monolithic or hybrid approaches and be scalable as well as affordable in fabrication \cite{Bogdanov2017}.
The silicon platform and the maturity of its fabrication make it a natural choice for integration.
However, for optical applications in communication, silicon has multiple drawbacks, including the lack of a direct bandgap, low transmission across the telecom C-band at high powers due to two-photon absorption and the lack of a second-order nonlinear coefficient.
These limitations make silicon a difficult choice for quantum communication systems using entangled photons \cite{OBrien2009,Wang2020}.

The ideal material parameters for an integrated quantum optical device are a direct bandgap, a high $\chi^{(2)}$ nonlinear coefficient, high index contrast, and low-loss waveguiding in the telecom C-band.
Commonly used material platforms are silicon nitride, which, except for the transparency window, inherits most of the drawbacks of silicon, lithium niobate and KTP, which lack the potential for active components, and indium phosphide and gallium arsenide, which suffer from losses due to a bandgap around 900\,nm wavelength.
A material platform without these drawbacks is aluminum gallium arsenide (AlGaAs), which has a large transparency window in the telecom wavelength range, integration capabilities for electro-optical devices like light-emitting diodes, lasers and modulators as well as a high second order nonlinear coefficient \cite{Shoji2002}.

In the form of Bragg-reflection waveguides (BRWs), AlGaAs simultaneously offers waveguiding and nonlinear conversion via modal phase matching \cite{Yeh1976,Helmy2006,Pressl2018}.
This enables the production of polarization, energy-time, and time-bin entangled photon pairs in the telecom wavelength range \cite{Horn2013,Valles2013,Schlager2017,Autebert2016,Chen2018}.
Additionally, owing to the direct bandgap of AlGaAs, a pump laser for nonlinear conversion processes or quantum dots for the creation of single photons can be directly integrated on chip \cite{Bijlani2013,Schlager2021,Boitier2014,Joens2015}.
As such, AlGaAs BRWs offer an ideal resource for quantum communication schemes in existing fiber networks.

While low loss is important in classical optical communication, it is paramount in entanglement-based quantum applications, as both photons of an entangled pair need to be preserved.
The fabrication of both classical AlGaAs devices, like lasers, photodetectors and modulators, as well as more complex quantum devices has seen significant advancements and has reached the level of sophistication required for large scale integration \cite{Porkolab2014,Volatier2010,Jiang2020,Chang2020}.

In this article, we report on the fabrication of low-loss AlGaAs BRWs, which serve as sources of correlated photon pairs.
By optimizing the etching recipe for near-vertical waveguide sidewalls and using resist reflow, we reduce the root-mean-square area sidewall roughness of our BRWs from 16.05(2)\,nm to 4.736(7)\,nm and the corresponding optical loss coefficient from between \( \alpha_\mathrm{no\,reflow}=0.23\left(9\right)\,\mathrm{mm}^{-1}\,\,\mathrm{and}\,\,0.32\left(9\right)\,\mathrm{mm}^{-1}\) to \( \alpha_\mathrm{reflow}=0.08\left(6\right)\,\mathrm{mm}^{-1}\).
Resist reflow increases the photon pair production rate by around six-fold.
Overall, the optimized fabrication recipe leads to a 15-fold increase in photon pair production rate compared to previous samples.

The rest of this paper is organized as follows.
We first introduce BRWs and their layer structure needed for mode guiding and phase matching.
The second part of the article details the fabrication steps and highlights the measures taken to handle the challenges introduced by the layer stack.
The third part focuses on the characterization of BRWs including measurements of the sidewall roughness and the loss coefficients via Fourier transforming the Fabry-Perot transmission spectrum \cite{Pressl2015}.
Finally, we demonstrate the increased correlated photon pair production rates.

\section{Bragg-reflection waveguides}

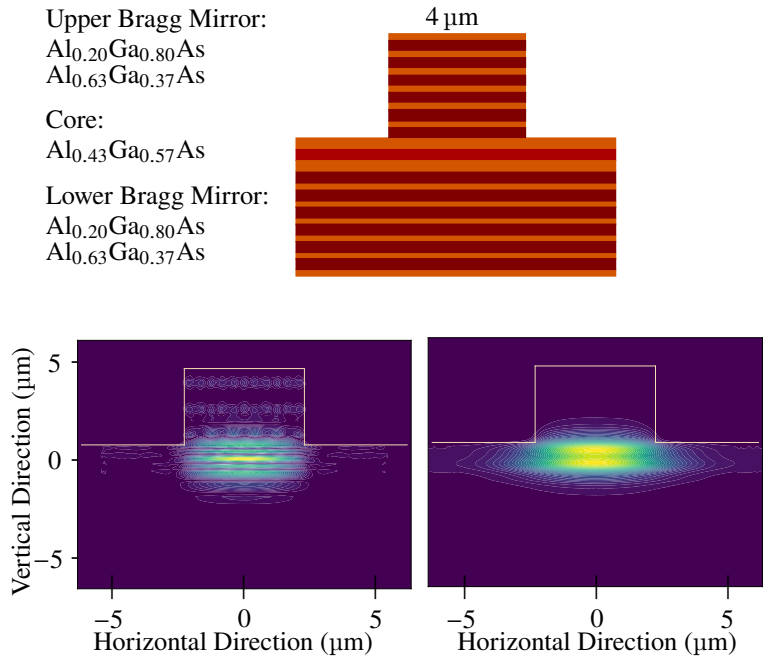
\begin{figure}[!htbp]
	\captionsetup{singlelinecheck = false, justification=raggedright}
    \begin{subfigure}[t]{0.62\textwidth}
    	 \centering
    	 \begin{annotate}{\includegraphics[width=1\textwidth]{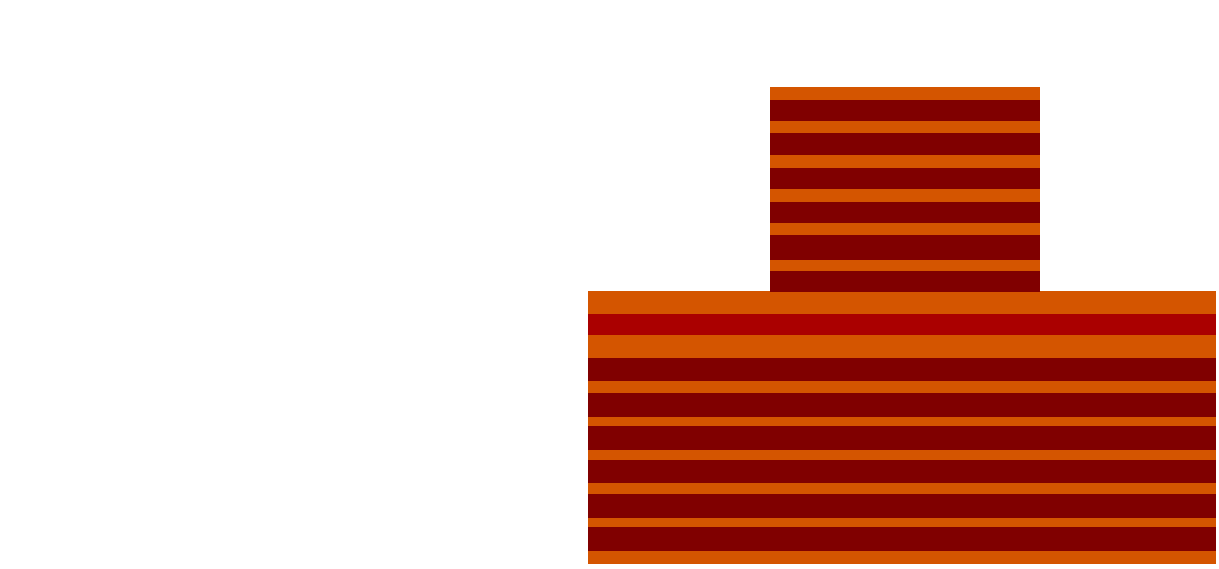}}{1}
    		\draw[very thick,black] (1.45,1.82) node[anchor=north west]{\SI{4}{\micro\meter}};
    		\draw[very thick,black] (-3.6,1.8) node[anchor=north west]{Upper Bragg Mirror:};
    		\draw[very thick,black] (-3.6,1.4) node[anchor=north west]{Al$_{0.20}$Ga$_{0.80}$As};
    		\draw[very thick,black] (-3.6,1.05) node[anchor=north west]{Al$_{0.63}$Ga$_{0.37}$As};
    		\draw[very thick,black] (-3.6,0.45) node[anchor=north west]{Core:};
    		\draw[very thick,black] (-3.6,0.05) node[anchor=north west]{Al$_{0.43}$Ga$_{0.57}$As};
    		\draw[very thick,black] (-3.6,-0.55) node[anchor=north west]{Lower Bragg Mirror:};
    		\draw[very thick,black] (-3.6,-0.95) node[anchor=north west]{Al$_{0.20}$Ga$_{0.80}$As};
    		\draw[very thick,black] (-3.6,-1.3) node[anchor=north west]{Al$_{0.63}$Ga$_{0.37}$As};
    	 \end{annotate}   
     \end{subfigure}
     
     \begin{subfigure}[t]{0.415\textwidth}
         \centering
         \begin{annotate}{\includegraphics[width=1\textwidth]{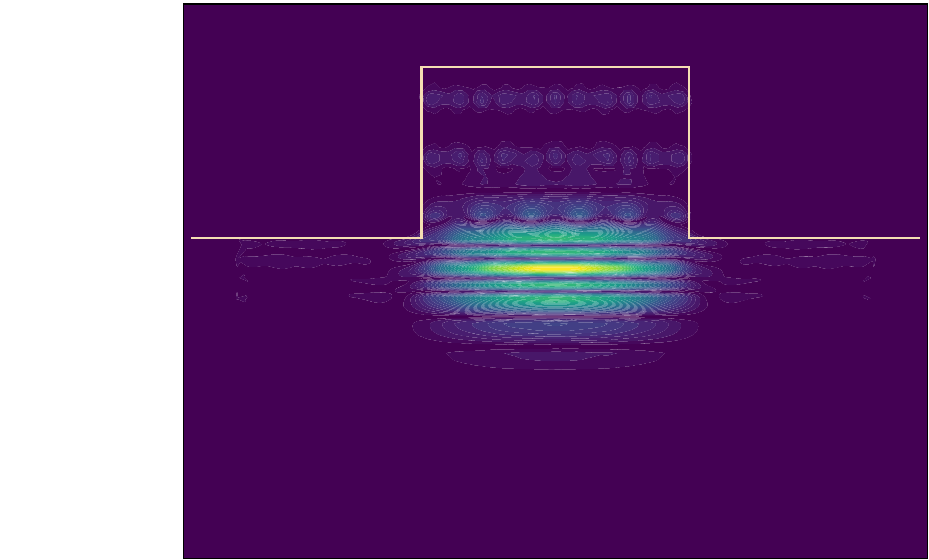}}{1}
    		\draw[very thick,black] (-1.6,-2.1) node[anchor=north west]{Horizontal Direction (\SI{}{\micro\meter})};
    		\draw[very thick,black] (-2.1,-1.9) node[anchor=south west, rotate=90]{Vertical Direction (\SI{}{\micro\meter})};
    		\draw[very thick,black] (-1.9,-1.1) node[anchor=north west]{-};
    		\draw[very thick,black] (-1.9,0.20) node[anchor=north west]{-};
    		\draw[very thick,black] (-1.9,1.50) node[anchor=north west]{-};
    		\draw[very thick,black] (-2.4,-1) node[anchor=north west]{$-5$};
    		\draw[very thick,black] (-2.2,0.30) node[anchor=north west]{$0$};
    		\draw[very thick,black] (-2.2,1.6) node[anchor=north west]{$5$};
    		\draw[very thick,black] (-1.4,-1.35) node[anchor=north west]{$\mid$};
    		\draw[very thick,black] (0.35,-1.35) node[anchor=north west]{$\mid$};
    		\draw[very thick,black] (2.1,-1.35) node[anchor=north west]{$\mid$};
    		\draw[very thick,black] (-1.6,-1.8) node[anchor=north west]{$-5$};
    		\draw[very thick,black] (0.32,-1.8) node[anchor=north west]{$0$};
    		\draw[very thick,black] (2.07,-1.8) node[anchor=north west]{$5$};
    	 \end{annotate}
     \end{subfigure}
     \begin{subfigure}[t]{0.35\textwidth}
         \begin{annotate}{\includegraphics[width=\textwidth]{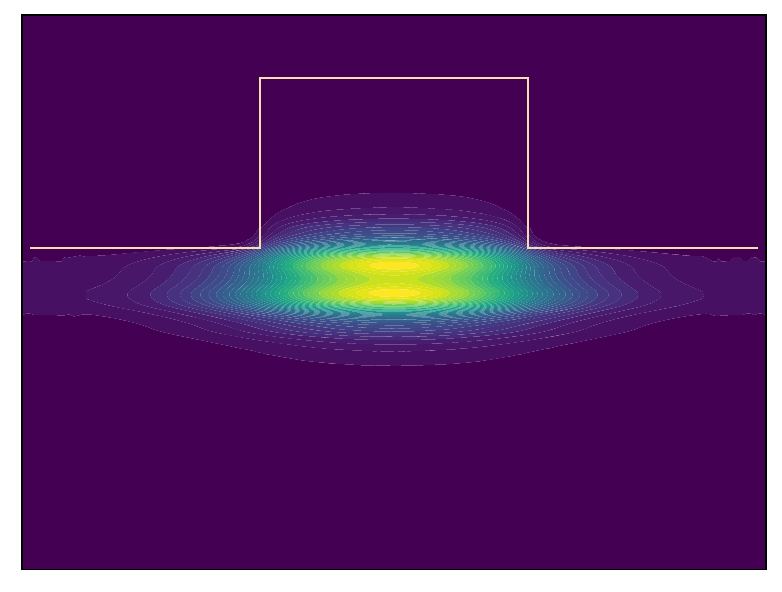}}{1}
    		\draw[very thick,black] (-1.95,-2.1) node[anchor=north west]{Horizontal Direction (\SI{}{\micro\meter})};
    		\draw[very thick,black] (-1.9,-1.35) node[anchor=north west]{$\mid$};
    		\draw[very thick,black] (-0.15,-1.35) node[anchor=north west]{$\mid$};
    		\draw[very thick,black] (1.6,-1.35) node[anchor=north west]{$\mid$};
    		\draw[very thick,black] (-2.1,-1.8) node[anchor=north west]{$-5$};
    		\draw[very thick,black] (-0.18,-1.8) node[anchor=north west]{$0$};
    		\draw[very thick,black] (1.55,-1.8) node[anchor=north west]{$5$};
    	 \end{annotate}    
     \end{subfigure}
        \caption{Top: Schematic of a BRW with layers of different aluminum concentration. The core (bright red) with a composition of Al$_{0.43}$Ga$_{0.57}$As lies between two optical matching layers and six Bragg mirror pairs above and below the core which are made up of Al$_{0.20}$Ga$_{0.80}$As (orange) and Al$_{0.63}$Ga$_{0.37}$As (dark red). Bottom: Simulations of the electric field absolute value for the TE-polarized Bragg mode at 775\,nm wavelength (left) and the total internal reflection mode at 1550\,nm wavelength for TE polarization (right). The image for the TM polarization at the latter wavelength is not shown as it is very similar to the TE one.}
    \label{fig:BRWLayers}
\end{figure}

The AlGaAs material platform features a high $\chi^{(2)}$ coefficient that allows nonlinear processes like second harmonic generation, difference-frequency generation, and parametric down-conversion (PDC).
The efficiency of these processes hinges on the momentum conservation between input and output photons as well as the low-loss guiding of the respective modes in the material.
We achieve low loss and phase matching in a BRW made of AlGaAs layers with varying aluminum concentrations.
Fig.~\ref{fig:BRWLayers} shows the layer structure and guided modes in a BRW.

A high index core realized via a low aluminum fraction guides two fundamental modes, one TE, the other TM polarized, at 1550\,nm wavelength via total internal reflection.
Two Bragg mirror stacks above and below the waveguide core confine the so-called Bragg mode at 775\,nm wavelength in the vertical direction.
These consist of six mirror pairs each.
We optimize the confinement by sizing the layer thicknesses at one quarter of the wavelength of the transverse component of the electric field at 775\,nm and by maximizing the difference in refractive index between the layers \cite{Yeh1976}.
The ridge structure of the waveguide ensures the confinement of the modes in the horizontal direction.
The BRW supports multiple spatial modes and the Bragg mode at 775\,nm wavelength is actually a higher order spatial mode.

Adjusting the refractive indices and layer thicknesses of the AlGaAs layers enables modal phase matching between the NIR Bragg and telecom fundamental modes of the waveguide.
For optimization, we take into account the increased nonlinear conversion efficiency with decreasing aluminum fraction as well as the simultaneous decrease in band gap and resulting pump absorption.
For a comprehensive guide to BRW design, refer to Ref. \cite{Pressl2018}.
While the elaborate layer structure of the BRWs affords wavelength flexibility and phase matching optimization in nonlinear processes, it poses a challenge for fabrication.

\section{Fabrication}\label{SecFab}

\begin{figure}[!htbp]
	\captionsetup{singlelinecheck = false, justification=raggedright}
	\centering
	\begin{annotate}{\includegraphics[width=3in]{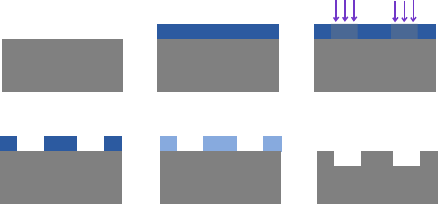}}{1}
    	\draw[very thick,black] (-3.9,1.85) node[anchor=north west]{a)};
    	\draw[very thick,black] (-1.15,1.85) node[anchor=north west]{b)};
    	\draw[very thick,black] (1.55,1.85) node[anchor=north west]{c)};
    	\draw[very thick,black] (-3.9,-0.1) node[anchor=north west]{d)};
    	\draw[very thick,black] (-1.15,-0.1) node[anchor=north west]{e)};
    	\draw[very thick,black] (1.55,-0.1) node[anchor=north west]{f)};
    	\draw[very thick,black] (-1.65,0.75) node[anchor=north west]{$\rightarrow$};
    	\draw[very thick,black] (1.1,0.75) node[anchor=north west]{$\rightarrow$};
    	\draw[very thick,black] (-1.65,-1.15) node[anchor=north west]{$\rightarrow$};
    	\draw[very thick,black] (1.1,-1.15) node[anchor=north west]{$\rightarrow$};
    	\draw[very thick,black] (-3.5,0.75) node[anchor=north west]{AlGaAs};
    	\draw[very thick,white] (-0.65,1.47) node[anchor=north west]{Resist};
    	\draw[very thick,black] (-0.75,0.75) node[anchor=north west]{AlGaAs};
    	\draw[very thick,black] (2,0.75) node[anchor=north west]{AlGaAs};
    	\draw[very thick,black] (-3.5,-1.15) node[anchor=north west]{AlGaAs};
    	\draw[very thick,black] (-0.75,-1.15) node[anchor=north west]{AlGaAs};
    	\draw[very thick,black] (2,-1.15) node[anchor=north west]{AlGaAs};
  	\end{annotate}
    \caption{BRW fabrication steps include oxygen plasma cleaning of the AlGaAs wafer surface (a), spinning of photoresist (b), photolithographic exposure (c), development (d), photoresist reflow (e), and etching (f).}
    \label{fig:FabSteps}
\end{figure}
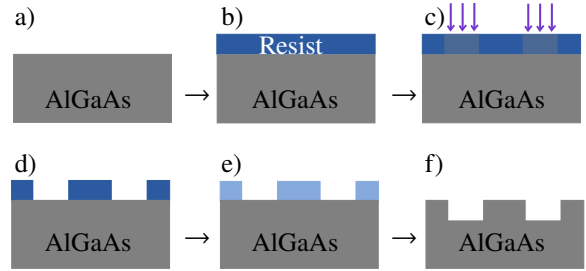

In the following section, three different fabrication recipes are described.
One that we call the ``previous" recipe uses e-beam lithography with stitching of write fields and metal mask deposition.
It is our starting point for improving the fabrication process of BRWs.
The new recipe is detailed in this section and is used in two versions: with and without reflowing the photoresist.
The epitaxial layers we use to fabricate BRWs are grown via molecular beam epitaxy on (100) GaAs substrates and are subsequently processed as shown in Fig.~\ref{fig:FabSteps}.
We first clean the wafer surface in an oxygen plasma in an inductively coupled plasma reactive ion etcher (ICP-RIE) (Sentech SI 500), the same machine that is later used for etching the waveguides.

A photo lithography machine (Microtech LaserWriter) exposes the waveguide pattern into the photoresist.
Using a direct laser writer and etch resistant resist allows us to readily adapt our design to different requirements.
Another advantage of this method is that it does not require an e-beam lithography machine and the more problematic chemicals needed for the deposition of a metal mask and lift-off.
Due to diffraction, the 405\,nm wavelength of the UV laser writer limits the feature size.
This, however, does not pose any problem for straight waveguides.
The lithography mode we employ is fixed-beam-moving-stage (FBMS) to avoid stitching errors between write fields.
This is the first step towards achieving smooth sidewalls and decreasing losses due to scattering.

\begin{figure}[!htbp]
	\captionsetup{singlelinecheck = false, justification=raggedright}
	\centering
	\begin{annotate}{\includegraphics[width=3in]{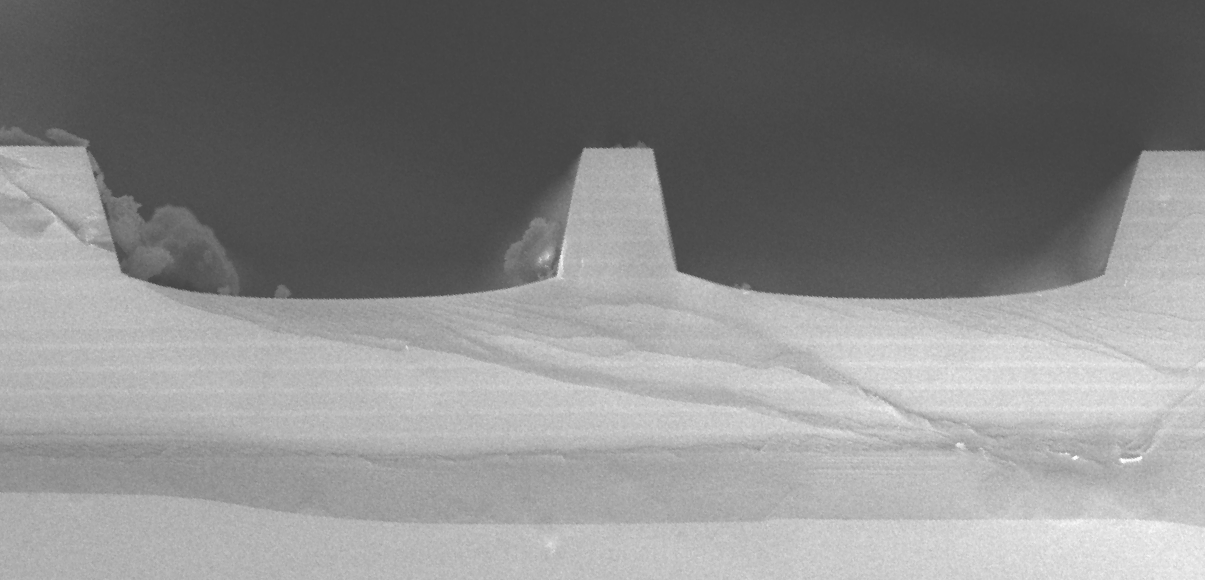}}{1}
		\draw[ultra thick,white] (-3.2,-0.84) node[anchor=north west]{------};
		\draw[ultra thick,white] (-3.2,-0.85) node[anchor=north west]{------};
		\draw[ultra thick,white] (-3.2,-0.86) node[anchor=north west]{------};
		\draw[ultra thick,white] (-3.2,-1) node[anchor=north west]{\SI{3}{\micro\meter}};
  	\end{annotate}
	
	\vspace{0.3cm}
	\begin{annotate}{\includegraphics[width=3in]{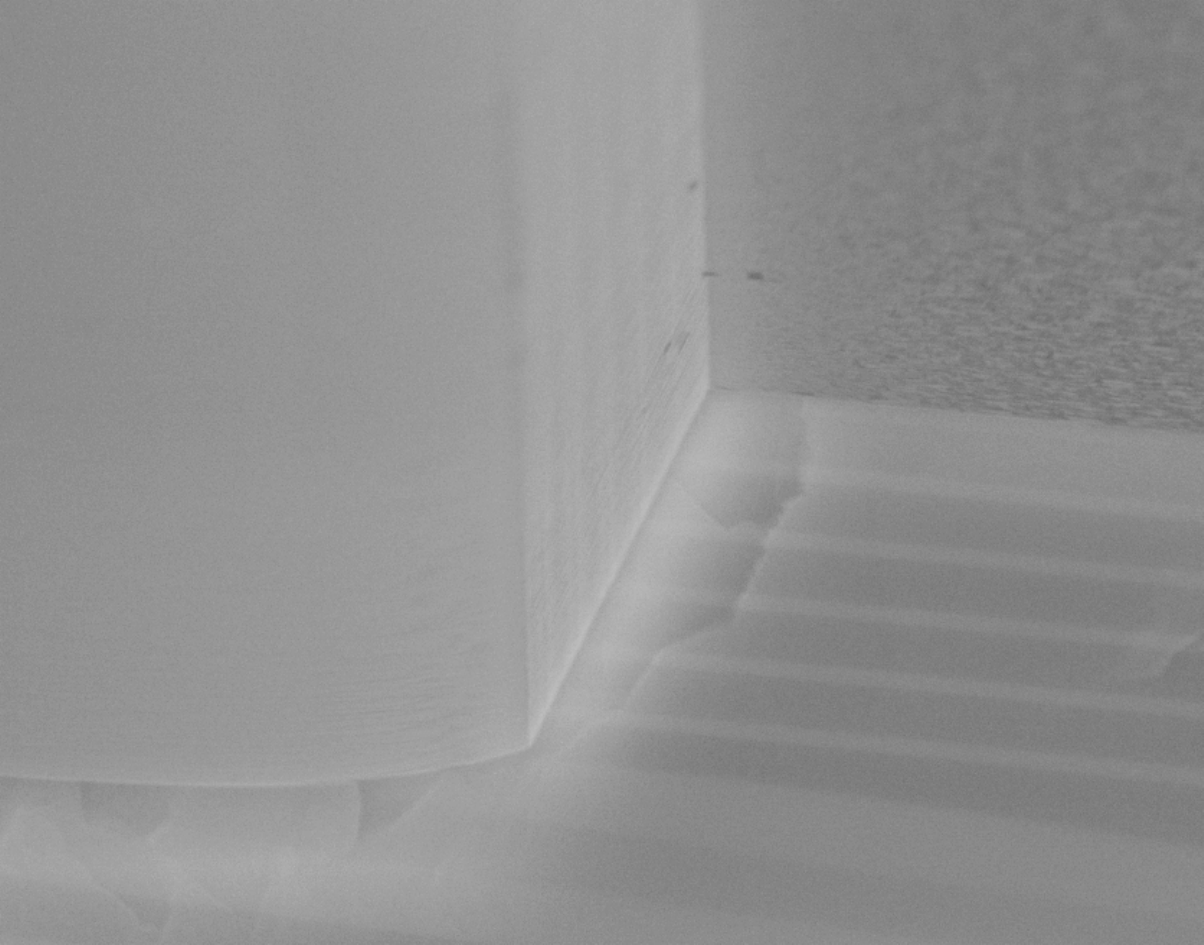}}{1}
		\draw[ultra thick,white] (-0.8,1.7) node[anchor=north west]{$\searrow$};
		\draw[ultra thick,white] (-1.8,2.1) node[anchor=north west]{Sidewall};
		\draw[ultra thick,white] (1.5,-0.9) node[anchor=north west]{Facet};
		\draw[ultra thick,white] (-3.2,-1.94) node[anchor=north west]{------};
		\draw[ultra thick,white] (-3.2,-1.95) node[anchor=north west]{------};
		\draw[ultra thick,white] (-3.2,-1.96) node[anchor=north west]{------};
		\draw[ultra thick,white] (-3.2,-2.1) node[anchor=north west]{\SI{1}{\micro\meter}};
  	\end{annotate}
    \caption{BRWs with homogeneously etched sidewalls in argon-borontrichloride plasma. A waveguide is defined by a trench on either side (top). The close-up of the sidewall shows the smooth corner between waveguide facet and sidewall (bottom).}
    \label{fig:BClEtch}
\end{figure}
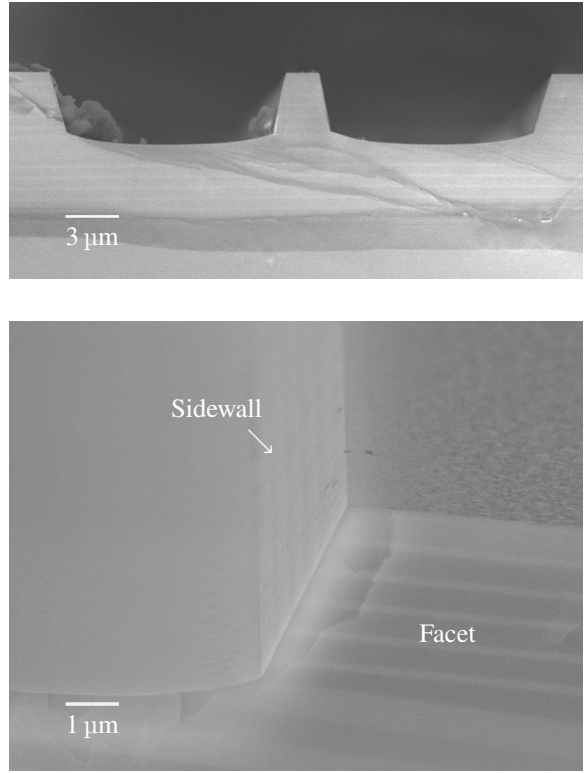

We spin the plasma-resistant photoresist AR-P3740 (Allresist) at 4000\,rpm to obtain a \SI{1.4}{\micro\meter} thick layer.
This allows us to etch the \SI{3.34}{\micro\meter} trench depth required for the BRWs with sufficient resist left to protect the surface of the wafer.
The photoresist is positive, enabling us to realize the structure shown in Fig.~\ref{fig:BClEtch} (top), which facilitates integration with other components and systems.
Each waveguide is defined by trenches etched on either side with most of the material between waveguides remaining intact.
This results in a shorter etch time, as less material needs to be removed, and more stable plasma, as less of the etched-away material accumulates in the etching chamber.
Removing less material during the etch also means that the chip can be handled more easily in semi-automatic assembly and flip-chip coupling.

After development of the photoresist in AR~300-35 (Allresist), we reflow the resist in a convection oven for 25~minutes at 140°C.
As the fluidity of the resist increases, its surface smoothens leading to a significant reduction in line edge roughness.
Any imperfections present in the photoresist transfer directly to the waveguide sidewall during the etching step and are evidenced by vertical striations.
Therefore, the reflow step is the most important in reducing sidewall roughness of the waveguides.

After reflow, we transfer the pattern in the resist to the AlGaAs in the ICP-RIE in a plasma mixture of argon and chlorine species.
The etch recipe needs to be finely tuned to achieve near-vertical, smooth sidewalls as well as non-selective etching in the lateral direction with respect to layers of different compositions.

\begin{figure}[!htbp]
	\captionsetup{singlelinecheck = false, justification=raggedright}
	\centering
	\begin{annotate}{\includegraphics[width=3in]{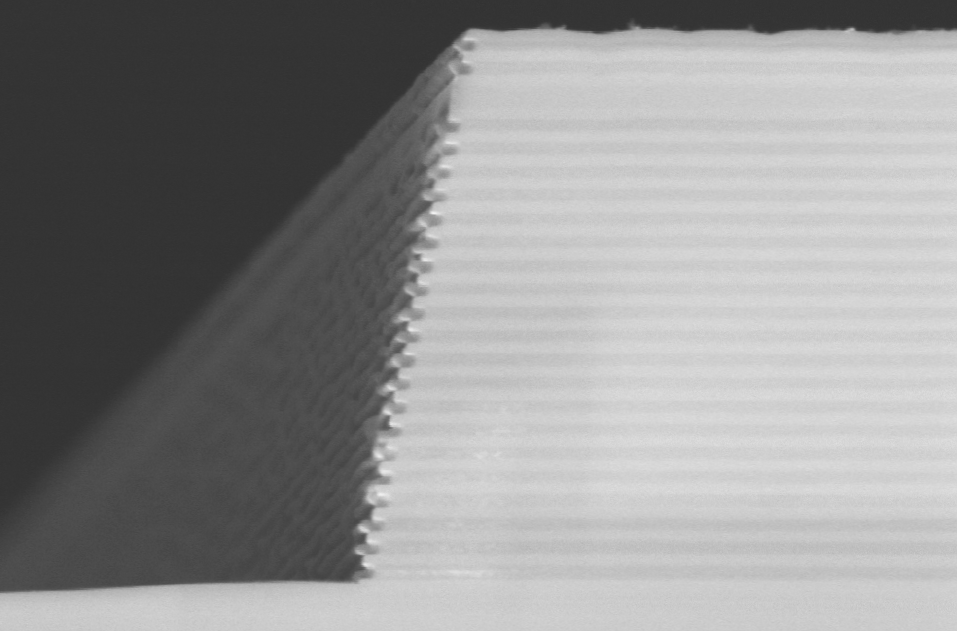}}{1}
		\draw[ultra thick,white] (-2,0.8) node[anchor=north west]{$\searrow$};
		\draw[ultra thick,white] (-3.2,1.2) node[anchor=north west]{Sidewall};
		\draw[ultra thick,black] (1,1.2) node[anchor=north west]{Facet};
		\draw[ultra thick,white] (-3.2,-1.47) node[anchor=north west]{------};
		\draw[ultra thick,white] (-3.2,-1.48) node[anchor=north west]{------};
		\draw[ultra thick,white] (-3.2,-1.46) node[anchor=north west]{------};
		\draw[ultra thick,white] (-3.2,-1.6) node[anchor=north west]{\SI{0.5}{\micro\meter}};
  	\end{annotate}
    \caption{Inhomogeneous lateral etch of layers with different aluminum fraction in high chlorine concentration recipes. Note that this is not a BRW sample but the aluminum fractions in the layers are comparable to BRW wafers.}
    \label{fig:ChlorineEtch}
\end{figure}
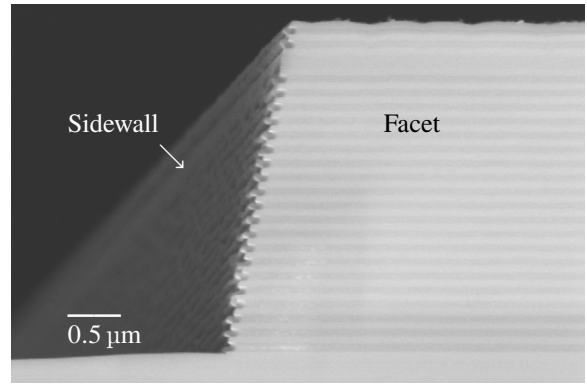

Vertical sidewalls are usually obtained in predominantly chemical etch processes where material is removed mainly via reactions with radicals in the plasma.
As radicals gather on the surface of the sample, etching occurs in the lateral direction removing material from the base of a waveguide ridge.
While such a chemical etch is fast and results in vertical sidewalls, the major drawback is the anisotropy of the etch at the sidewalls.
Layers with lower aluminum concentrations etch faster than those with a higher aluminum fraction, as shown in Fig.~\ref{fig:ChlorineEtch}.
The aluminum, having a lower electronegativity than the gallium, tends to react with other components available in the plasma.
This could be resist material that has been removed from the surface by the etch gases.
The resulting aluminum oxide layers then protect the high aluminum content layers from further chemical etching \cite{Bruce1983}.
This roughness is not homogeneous along the length of the waveguide and thus contributes immensely to scattering losses.

Oxidation films on high aluminum fraction layers can best be avoided in physical etch processes.
Here, charged atoms and molecules impinging on the sample surface remove material.
This role is mainly played by the argon atoms in our etch recipe.
This sputter component is increased by replacing the chlorine with boron trichloride.
Boron trichloride undergoes dissociation and recombination processes in the etch chamber and thereby reduces the amount of free chlorine available for the etch.
In addition, it reacts with aluminum oxide and water vapor lessening the inhomogeneity of the sidewalls~\cite{Bruce1983}.

The chemical mixture of the etch gas alone, however, is not enough to attain vertical and non-selectively etched sidewalls at the same time.
We therefore tune the other settings of the ICP-RIE. We flood the etch chamber with 40\,SCCM of argon and 4\,SCCM of boron trichloride at a low pressure of 0.3\,Pa.
While the effects of the pressure on the etch strongly depend on all other parameters, we find in this regime that the low pressure avoids the accumulation of reactive species on the sample surface.

The plasma is ignited via an RF-coupled induction coil at the top of the etch chamber.
The ICP power is set to a commonly employed value of 400\,W.
In order to accelerate the ions in the plasma towards the sample, we apply an RF signal that leads to a self-biasing of the sample holder.
We obtain high-momentum particles impinging on the sample surface anisotropically by setting the self-bias to a high value of 150\,V.
This makes the etch sufficiently physical for the removal of oxidation products, leading to homogeneous sidewalls.
It also means that the ions reach the lower parts of the already etched sidewall, leading to an almost vertical sidewall otherwise not possible in such a physical etch.
Two scanning electron micrographs of a BRW fabricated in this way are shown in Fig.~\ref{fig:BClEtch}.
The SEM images show no noticeable inhomogeneity between the different layers on the sidewall.
Small vertical striations are visible and will be analyzed in the following section.

\section{Characterization}

We assess the quality of the BRWs by determining the sidewall roughness, optical loss coefficient, and photon pair production rate in PDC.
In order to quantify the improvement of the fabrication after adding the reflow step into the recipe, we employ two different methods.
We first determine the sidewall roughness using an atomic force microscope (AFM) and then measure the optical loss in the BRW via a Fourier transform analysis of the transmission spectrum for the wavelength range around 1550\,nm.

Assessment of the sidewall roughness is done using an AFM (Nanosurf NaioAFM) with a \(<10\,\mathrm{nm}\) radius cantilever tip (BudgetSensors Tap190GD-G).
As we would like to separate the effect of the reflow from the inhomogeneity introduced by the layer stack, we perform this analysis on GaAs waveguides etched in the same recipe as the BRWs.
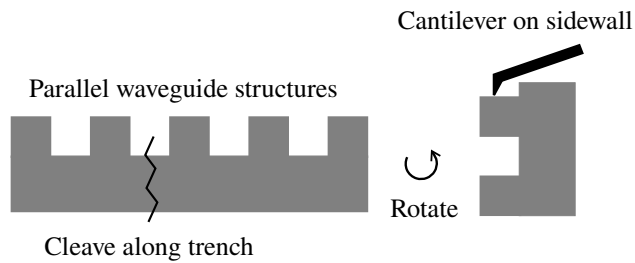
\begin{figure}[!htbp]
	\captionsetup{singlelinecheck = false, justification=raggedright}
	\centering
	\begin{annotate}{\includegraphics[width=3in]{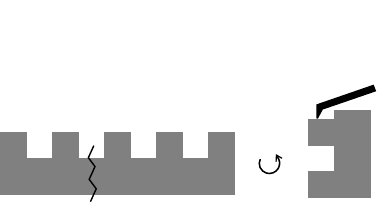}}{1}
    	\draw[very thick,black] (-3.7,0) node[anchor=north west]{Parallel waveguide structures};
    	\draw[very thick,black] (-3.5,-2.1) node[anchor=north west]{Cleave along trench};
    	\draw[very thick,black] (1.1,-1.6) node[anchor=north west]{Rotate};
    	\draw[very thick,black] (1.2,0.9) node[anchor=north west]{Cantilever on sidewall};
  	\end{annotate}
    \caption{After cleaving along a trench between waveguides, we rotate our sample such that the AFM tip has access to the full width and height of the waveguide's sidewall.}
    \label{fig:AFMTechnique}
\end{figure}

\begin{figure}[!htbp]
	\captionsetup{singlelinecheck = false, justification=raggedright}
	\centering
	\begin{annotate}{\includegraphics[width=3.3in]{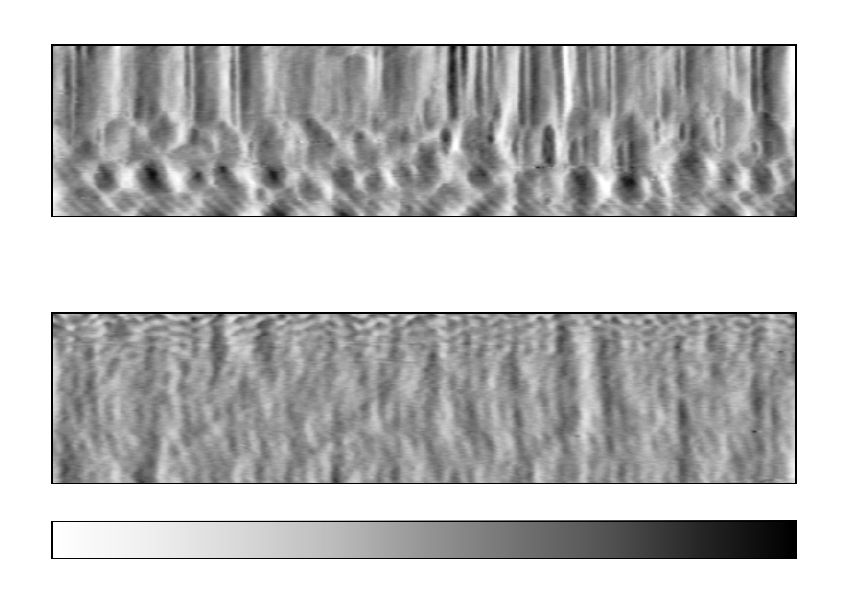}}{1}
    	\draw[very thick,black] (-1.3,0.5) node[anchor=north west]{Image Length (\SI{}{\micro\meter})};
		\draw[very thick,black] (-1.2,-3) node[anchor=north west]{Pixel Height (nm)};
		\draw[very thick,black] (-4.5,-1.5) node[anchor=north west, rotate=90]{Image Width (\SI{}{\micro\meter})};
		\draw[very thick,black] (-4.2,-2.62) node[anchor=north west]{-10.0};
		\draw[very thick,black] (-3.16,-2.62) node[anchor=north west]{-7.5};
		\draw[very thick,black] (-2.22,-2.62) node[anchor=north west]{-5.0};
		\draw[very thick,black] (-1.28,-2.62) node[anchor=north west]{-2.5};
		\draw[very thick,black] (-0.3,-2.62) node[anchor=north west]{0.0};
		\draw[very thick,black] (0.65,-2.62) node[anchor=north west]{2.5};
		\draw[very thick,black] (1.58,-2.62) node[anchor=north west]{5.0};
		\draw[very thick,black] (2.50,-2.62) node[anchor=north west]{7.5};
		\draw[very thick,black] (3.35,-2.62) node[anchor=north west]{10.0};
		\draw[very thick,black] (-3.87,0.77) node[anchor=north west]{0};
		\draw[very thick,black] (-2.92,0.77) node[anchor=north west]{1};
		\draw[very thick,black] (-1.97,0.77) node[anchor=north west]{2};
		\draw[very thick,black] (-1.02,0.77) node[anchor=north west]{3};
		\draw[very thick,black] (-0.07,0.77) node[anchor=north west]{4};
		\draw[very thick,black] (0.88,0.77) node[anchor=north west]{5};
		\draw[very thick,black] (1.83,0.77) node[anchor=north west]{6};
		\draw[very thick,black] (2.78,0.77) node[anchor=north west]{7};
		\draw[very thick,black] (-4.14,1.07) node[anchor=north west]{0};
		\draw[very thick,black] (-4.14,1.93) node[anchor=north west]{1};
		\draw[very thick,black] (-4.14,-1.55) node[anchor=north west]{0};
		\draw[very thick,black] (-4.14,-0.69) node[anchor=north west]{1};
		\draw[very thick,black] (-3.94,1.9) node[anchor=north west]{--};
		\draw[very thick,black] (-3.94,1.03) node[anchor=north west]{--};
		\draw[very thick,black] (-3.94,-1.63) node[anchor=north west]{--};
		\draw[very thick,black] (-3.94,-0.76) node[anchor=north west]{--};
		\draw[very thick,black] (-3.85,0.58) node[anchor=north west, rotate=90]{--};
		\draw[very thick,black] (-2.9,0.58) node[anchor=north west, rotate=90]{--};
		\draw[very thick,black] (-1.95,0.58) node[anchor=north west, rotate=90]{--};
		\draw[very thick,black] (-1,0.58) node[anchor=north west, rotate=90]{--};
		\draw[very thick,black] (-0.05,0.58) node[anchor=north west, rotate=90]{--};
		\draw[very thick,black] (0.9,0.58) node[anchor=north west, rotate=90]{--};
		\draw[very thick,black] (1.85,0.58) node[anchor=north west, rotate=90]{--};
		\draw[very thick,black] (2.8,0.58) node[anchor=north west, rotate=90]{--};
		\draw[very thick,black] (-3.85,-0.27) node[anchor=north west, rotate=90]{--};
		\draw[very thick,black] (-2.9,-0.27) node[anchor=north west, rotate=90]{--};
		\draw[very thick,black] (-1.95,-0.27) node[anchor=north west, rotate=90]{--};
		\draw[very thick,black] (-1,-0.27) node[anchor=north west, rotate=90]{--};
		\draw[very thick,black] (-0.05,-0.27) node[anchor=north west, rotate=90]{--};
		\draw[very thick,black] (0.9,-0.27) node[anchor=north west, rotate=90]{--};
		\draw[very thick,black] (1.85,-0.27) node[anchor=north west, rotate=90]{--};
		\draw[very thick,black] (2.8,-0.27) node[anchor=north west, rotate=90]{--};
		\draw[very thick,black] (-3.85,-2.82) node[anchor=north west, rotate=90]{--};
		\draw[very thick,black] (-2.9225,-2.82) node[anchor=north west, rotate=90]{--};
		\draw[very thick,black] (-1.995,-2.82) node[anchor=north west, rotate=90]{--};
		\draw[very thick,black] (-1.0675,-2.82) node[anchor=north west, rotate=90]{--};
		\draw[very thick,black] (-0.14,-2.82) node[anchor=north west, rotate=90]{--};
		\draw[very thick,black] (0.7875,-2.82) node[anchor=north west, rotate=90]{--};
		\draw[very thick,black] (1.715,-2.82) node[anchor=north west, rotate=90]{--};
		\draw[very thick,black] (2.6425,-2.82) node[anchor=north west, rotate=90]{--};
		\draw[very thick,black] (3.568,-2.82) node[anchor=north west, rotate=90]{--};
	\end{annotate}
    \caption{Cutouts of the full AFM images, with low Fourier components omitted. The sample made without the reflow step shows clear vertical striations near the top of the waveguide and a honeycomb-like structure towards the substrate (top). Reflowing the resist smoothens the sidewall (bottom).}
    \label{fig:Sidewalls}
\end{figure}
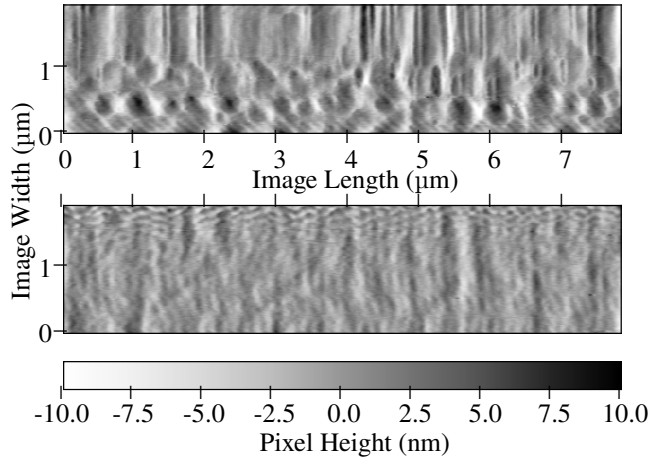

We cleave the sample along the trench on one side of the waveguide and then mount it in the AFM sample holder with the waveguide sidewall facing upward.
This offers access for the AFM tip to the entire length and width of the sidewall, as illustrated in Fig.~\ref{fig:AFMTechnique}.
We scan squares of \SI{8}{\micro\meter} width on the samples fabricated with and without reflow and compare them.
Fig.~\ref{fig:Sidewalls} shows cutouts of two such profiles corresponding to the middle parts of the sidewalls.
These can be reproducibly imaged by the AFM as opposed to the sections near the substrate or the upper edge of the sidewall.
In order to improve the illustration of the sidewall roughness in this figure, we enhance both images by removing the same low spatial frequency components from each horizontal line profile.
These low components result from the way the cantilever moves across the sample, moving slightly upward in the center of a scanned section.
For the quantitative investigation of the sidewall roughness, we use the original AFM data.
From the AFM profiles, we calculate the root-mean-square area roughness of the sidewalls

 \begin{equation}
	\sigma_\mathrm{RMS} = \sqrt{\frac{\sum _{\substack{
   N }}(z-\bar{z})^2}{N -1}}
	\label{eqn:AreaRoughness}
\end{equation}
 where \(N\) is the number of pixels, \(z\) is the height value recorded for a certain pixel and \(\bar{z}\) is the mean height.
We obtain values of 16.05(2)\,nm for the no-reflow samples and 4.736(7)\,nm for the samples fabricated using the reflow method.
Absolute area roughness values depend strongly on the cantilever tip and measurement mode, but the relative roughness can be estimated. 
Here, the area roughness is reduced by a factor of more than three when employing the reflow method.

From the findings in the AFM studies on GaAs, we expect the scattering losses in the reflow BRW sample to be substantially lower than in previous samples.
This is because we consider the sidewall roughness a variation in the waveguide width, and thus a grating coupler to the radiation modes \cite{Melati2014,Payne1994,Roberts2022,Jaberansary2013}.
The decreased scattering losses are confirmed by measuring the transmission spectrum of the BRW in the telecom wavelength range, where the photons are produced during PDC.
We couple a tunable telecom laser (Santec TSL-710) into the waveguide via a 100x microscope objective and collect the output light using an aspheric lens.
A powermeter (Thorlabs S122C) measures the transmitted light.
The facets of the waveguide have a reflectivity of 35(4)\%, meaning the waveguide acts as a weak cavity~\cite{Pressl2015}.
As light from multiple reflections interferes at the output facet of the waveguide, the power meter detects interference fringes when the wavelength of the input light is scanned.
In a single mode system, one could readily determine the losses in the waveguide from the visibility of these Fabry-Perot fringes.
The BRW, however, supports higher order spatial modes including the Bragg mode.
Due to the multimodal nature of the waveguide, the fringes exhibit a beat pattern that is more difficult to interpret.
In order to modally resolve this transmissivity measurement, we Fourier-transform the transmission spectrum as detailed in Ref. \cite{Pressl2015}.

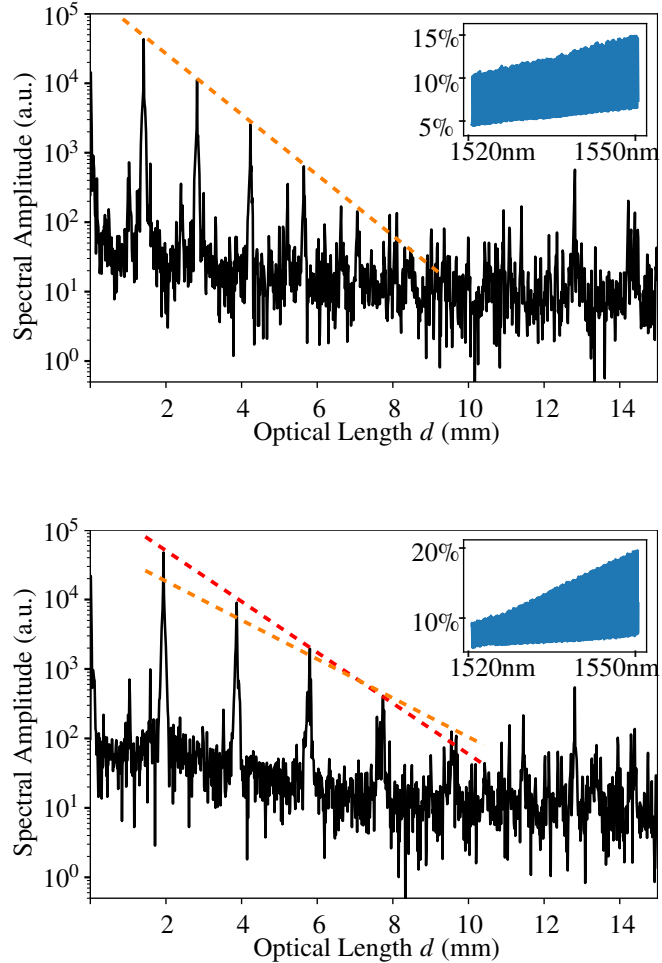
\begin{figure}[!htbp]
	\captionsetup{singlelinecheck = false, justification=raggedright}
	\centering
	\begin{annotate}{\includegraphics[width=3in]{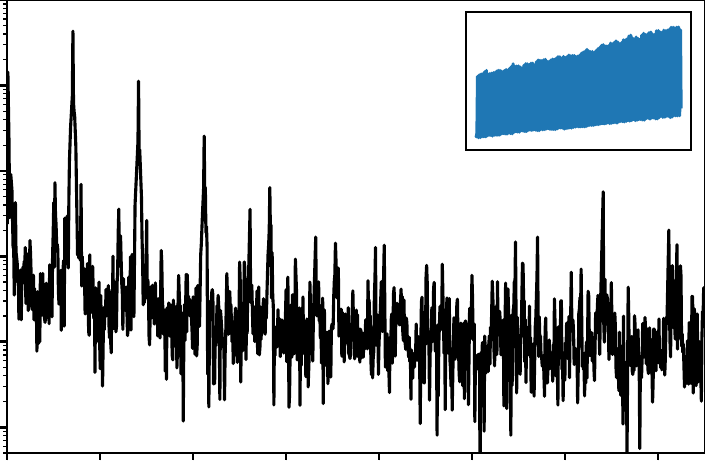}}{1}
		\draw[very thick,black] (-1.7,-2.8) node[anchor=north west]{Optical Length $d$ (mm)};
		\draw[very thick,black] (-2.93,-2.5) node[anchor=north west]{2};
		\draw[very thick,black] (-1.92,-2.5) node[anchor=north west]{4};
		\draw[very thick,black] (-0.91,-2.5) node[anchor=north west]{6};
		\draw[very thick,black] (0.09,-2.5) node[anchor=north west]{8};
		\draw[very thick,black] (1,-2.5) node[anchor=north west]{10};
		\draw[very thick,black] (2.02,-2.5) node[anchor=north west]{12};
		\draw[very thick,black] (3.03,-2.5) node[anchor=north west]{14};
		\draw[very thick,black] (-4.3,-2) node[anchor=south west, rotate=90]{Spectral Amplitude (a.u.)};
		\draw[very thick,black] (-4.5,-2.4) node[anchor=south west]{\(10^0\)};
		\draw[very thick,black] (-4.5,-1.47) node[anchor=south west]{\(10^1\)};
		\draw[very thick,black] (-4.5,-0.55) node[anchor=south west]{\(10^2\)};
		\draw[very thick,black] (-4.5,0.37) node[anchor=south west]{\(10^3\)};
		\draw[very thick,black] (-4.5,1.31) node[anchor=south west]{\(10^4\)};
		\draw[very thick,black] (-4.5,2.2) node[anchor=south west]{\(10^5\)};
		\draw[very thick,black] (0.9,0.4) node[anchor=south west]{1520nm};
		\draw[very thick,black] (2.55,0.4) node[anchor=south west]{1550nm};
		\draw[very thick,black] (0.5,0.72) node[anchor=south west]{5\(\%\)};
		\draw[very thick,black] (0.4,1.3) node[anchor=south west]{10\(\%\)};
		\draw[very thick,black] (0.4,1.9) node[anchor=south west]{15\(\%\)};
		\draw[very thick,black] (1.02,1.21) node[anchor=north west]{-};
		\draw[very thick,black] (1.02,1.78) node[anchor=north west]{-};
		\draw[very thick,black] (1.02,2.35) node[anchor=north west]{-};
		\draw[very thick,black] (1.15,0.65) node[anchor=north west, rotate=90]{-};
		\draw[very thick,black] (3.37,0.65) node[anchor=north west, rotate=90]{-};
		\draw [dashed][line width=0.5mm, orange ] (-3.3,2.42) -- (0.96,-1) node [midway] {};
  	\end{annotate}
	
	\vspace{0.7cm}
	\begin{annotate}{\includegraphics[width=3in]{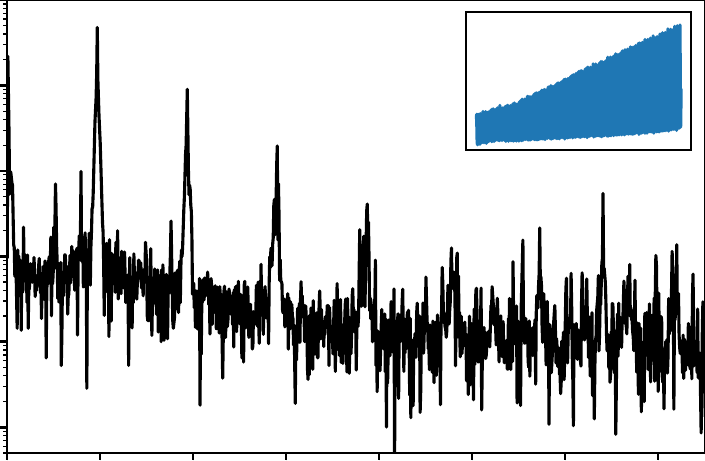}}{1}
		\draw[very thick,black] (-1.7,-2.8) node[anchor=north west]{Optical Length $d$ (mm)};
		\draw[very thick,black] (-2.93,-2.5) node[anchor=north west]{2};
		\draw[very thick,black] (-1.92,-2.5) node[anchor=north west]{4};
		\draw[very thick,black] (-0.91,-2.5) node[anchor=north west]{6};
		\draw[very thick,black] (0.09,-2.5) node[anchor=north west]{8};
		\draw[very thick,black] (1,-2.5) node[anchor=north west]{10};
		\draw[very thick,black] (2.02,-2.5) node[anchor=north west]{12};
		\draw[very thick,black] (3.03,-2.5) node[anchor=north west]{14};
		\draw[very thick,black] (-4.3,-2) node[anchor=south west, rotate=90]{Spectral Amplitude (a.u.)};
		\draw[very thick,black] (-4.5,-2.4) node[anchor=south west]{\(10^0\)};
		\draw[very thick,black] (-4.5,-1.47) node[anchor=south west]{\(10^1\)};
		\draw[very thick,black] (-4.5,-0.55) node[anchor=south west]{\(10^2\)};
		\draw[very thick,black] (-4.5,0.37) node[anchor=south west]{\(10^3\)};
		\draw[very thick,black] (-4.5,1.31) node[anchor=south west]{\(10^4\)};
		\draw[very thick,black] (-4.5,2.2) node[anchor=south west]{\(10^5\)};
		\draw[very thick,black] (0.9,0.4) node[anchor=south west]{1520nm};
		\draw[very thick,black] (2.55,0.4) node[anchor=south west]{1550nm};
		\draw[very thick,black] (0.4,1.0) node[anchor=south west]{10\(\%\)};
		\draw[very thick,black] (0.4,1.93) node[anchor=south west]{20\(\%\)};
		\draw[very thick,black] (1.02,1.47) node[anchor=north west]{-};
		\draw[very thick,black] (1.02,2.4) node[anchor=north west]{-};
		\draw[very thick,black] (1.15,0.65) node[anchor=north west, rotate=90]{-};
		\draw[very thick,black] (3.37,0.65) node[anchor=north west, rotate=90]{-};
		\draw [dashed][line width=0.5mm, red ] (-3,2.4) -- (1.5,-0.63) node [midway] {};
		\draw [dashed][line width=0.5mm, orange ] (-3,1.95) -- (1.5,-0.37) node [right] {};	
  	\end{annotate}
    \caption{Fourier transform of the transmission spectrum (inset) for TE polarized input light in the telecom wavelength range for the BRW fabricated without (top) and with reflow (bottom). For the latter, the peak height ratios of early peaks (red line) differ from those of later peaks (orange line). An early peak refers to a peak farther to the left in the plot and corresponds to light that has only survived one or two passes through the waveguide cavity. Light that contributes to a late peak farther to the right has passed through the waveguide four or five times.}
    \label{fig:FourierTransmission}
\end{figure}

The resulting Fourier spectra, shown in Fig.~\ref{fig:FourierTransmission}, can be interpreted as follows.
Light in the waveguide experiences a time delay at every pass through the waveguide and at every reflection from the facets.
This time delay corresponds to a translation in Fourier space and manifests as peaks at integer multiples of the resonator length for each mode.
The Fourier spectra shown here feature a pattern of one strong peak and smaller side peaks that repeat again after a distance of one resonator length.
This corresponds to one or more strong modes at similar effective refractive index and weaker modes with different effective refractive indices.
The ratio of heights \(\tilde{R}\) of subsequent peaks belonging to one mode thus contains information about the loss during one pass through the waveguide for that specific mode.
For a known reflectivity of the facets \(R\) and resonator length \(L\), the loss coefficient can be readily calculated:

\begin{equation}
 	\alpha = -\frac{1}{L} \mathrm{ln}\left(\frac{\tilde{R}}{R}\right).
	\label{eqn:LossCoeff}
\end{equation}

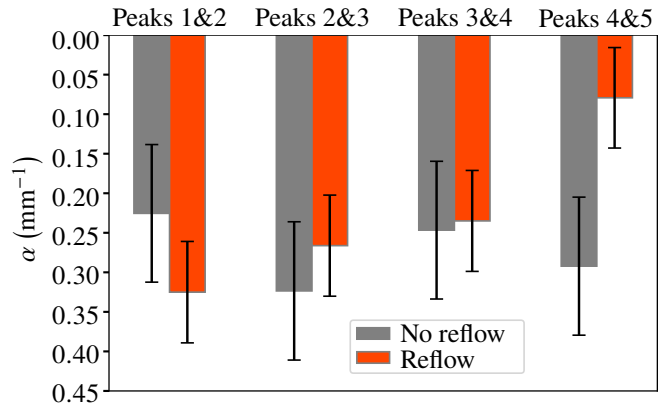
\begin{figure}[!htbp]
	\captionsetup{singlelinecheck = false, justification=raggedright}
	\centering
    \begin{annotate}{\includegraphics[width=2.9in]{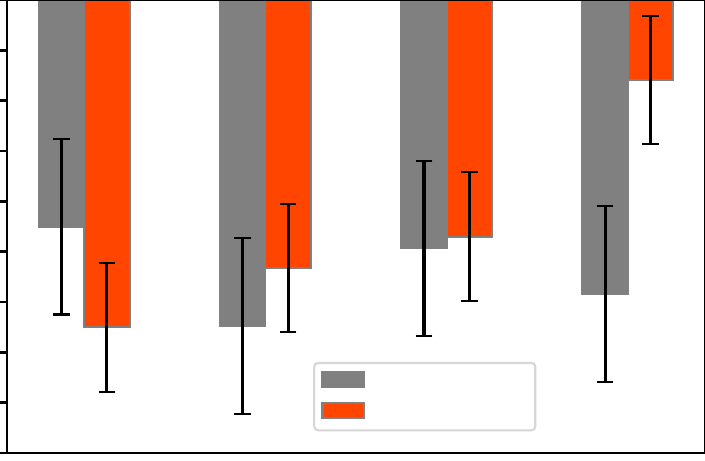}}{1}
    	\draw[very thick,black] (-3.7,2.9) node[anchor=north west]{Peaks \(1\&2\)};
    	\draw[very thick,black] (-1.8,2.9) node[anchor=north west]{Peaks \(2\&3\)};
    	\draw[very thick,black] (0.10,2.9) node[anchor=north west]{Peaks \(3\&4\)};
    	\draw[very thick,black] (1.97,2.9) node[anchor=north west]{Peaks \(4\&5\)};
    	\draw[very thick,black] (0.13,-1.3) node[anchor=north west]{No reflow};
    	\draw[very thick,black] (0.13,-1.65) node[anchor=north west]{Reflow};
    	\draw[very thick,black] (-4.35,-0.75) node[anchor=south west, rotate=90]{\(\alpha \left(\mathrm{mm}^{-1}\right)\)};
    	\draw[very thick,black] (-4.5,2.6) node[anchor=north west]{$0.00$};
    	\draw[very thick,black] (-4.5,2.078) node[anchor=north west]{$0.05$};
    	\draw[very thick,black] (-4.5,1.556) node[anchor=north west]{$0.10$};
    	\draw[very thick,black] (-4.5,1.034) node[anchor=north west]{$0.15$};
    	\draw[very thick,black] (-4.5,0.512) node[anchor=north west]{$0.20$};
    	\draw[very thick,black] (-4.5,-0.01) node[anchor=north west]{$0.25$};
    	\draw[very thick,black] (-4.5,-0.532) node[anchor=north west]{$0.30$};
    	\draw[very thick,black] (-4.5,-1.054) node[anchor=north west]{$0.35$};
    	\draw[very thick,black] (-4.5,-1.576) node[anchor=north west]{$0.40$};
    	\draw[very thick,black] (-4.5,-2.098) node[anchor=north west]{$0.45$};
    \end{annotate}   
    \caption{Loss coefficients $\alpha$ derived from the Fourier transform of the transmission spectrum of two BRWs fabricated with reflow shown in orange and those fabricated without in grey. The bars represent the loss coefficients calculated from the ratios of neighboring peaks in the Fourier spectrum.}
    \label{fig:Alphas}
\end{figure}

\noindent The optical length $d$ is proportional to $L$ via the factor $\pi/n$, where \(n\) is the group refractive index.
We determine the loss coefficient for each ratio of neighboring peak heights separately.
Fig.~\ref{fig:Alphas} shows a comparison of the loss coefficients of two BRWs made with (orange) and without (grey) the reflow step but coming from the same wafer.
The loss coefficient of the latter varies between \( \alpha_\mathrm{no\,reflow}=0.23\left(9\right)\,\mathrm{mm}^{-1}\) and \( \alpha_\mathrm{no\,reflow}=0.32\left(9\right)\,\mathrm{mm}^{-1}\) without any upward or downward trend.
The loss coefficient of the BRW fabricated with reflow is higher than that of the no-reflow sample for the first two peaks in the Fourier spectrum.
This is counterintuitive, as we would expect a smoother sidewall leading to lower loss coefficients.
However, for later peaks, the loss coefficient drops below that of the BRW made without reflowed resist.
It also follows a clear trend, decreasing down to a loss coefficient of \( \alpha_\mathrm{reflow}=0.08\left(6\right)\,\mathrm{mm}^{-1}\).

All but the fundamental telecom mode are strongly dampened within the waveguide, which was manufactured without reflow.
Mode simulations show that higher-order modes exhibit higher field strengths closer to the waveguide surface and are thus more easily scattered with increased surface roughness.
Thus, mainly the fundamental mode survives.
When the sidewall roughness is reduced by reflowing the resist, another, higher order mode with higher loss can also propagate within the waveguide.
This mode has an effective refractive index difference to the fundamental mode of only about 0.01 at 1535\(\,\)nm wavelength. 
This higher-loss mode dominates the early peak height ratios, causing high loss coefficients.
As its contribution diminishes, the losses go down, revealing how much the reflow step  improved the propagation of the original mode.
Overall, the losses for TE polarized light in the telecom wavelength range were reduced from between \( \alpha_\mathrm{no\,reflow}=0.23\left(9\right)\,\mathrm{mm}^{-1}\,\,\mathrm{and}\,\,0.32\left(9\right)\,\mathrm{mm}^{-1}\) to \( \alpha_\mathrm{reflow}=0.08\left(6\right)\,\mathrm{mm}^{-1}\).

Reducing the losses on this scale has a significant effect on nonlinear conversion processes.
We test this by measuring the coincidence rates between the signal and idler photons produced in PDC.
To do so, we pump the sample with a NIR laser (MSquared SolsTis 6W PSX-R) set to the phase-matching wavelength.
A dichroic mirror placed at the output of the BRW filters out the remaining pump light.
The down-converted photon pair is orthogonally polarized and can therefore be split up at a polarizing beam splitter and relayed to two superconducting nanowire single photon detectors (Single Quantum Eos 720 CS).
We measure the coincidence rate between the two detectors for a pump power of 1\,mW and normalize to the length of the waveguide sample. Table~\ref{tab:PDCData} lists the results for the BRWs fabricated using the recipe with and without reflow detailed above, and the previous recipe employing e-beam lithography and a metal mask for etching.
The coincidence rate and therefore the conversion efficiency is highest in the sample fabricated using the reflow step.
Both of the recipes employing FBMS optical lithography perform better than the recipe using e-beam lithography and metal mask deposition.
By adding the reflow step to the recipe, the coincidence rate could be increased by around six-fold.
Note that the measured coincidence rates depend on the coupling efficiencies in the setup.
The values shown here were measured at similar conditions in the setup to allow a comparison.
However, much higher coincidence rates of \(89\left(5\right)\,\mathrm{Hz/\mathrm{\mu} W}\) have been achieved in the meantime using the reflow-sample in an optimized setup \cite{Nardi2022}.
Such an improvement in performance through careful optimization of the fabrication recipe means that BRWs are a reliable option where an integrable source for photon pairs is needed.

\begin{table}[!htbp]
\captionsetup{singlelinecheck = false, justification=raggedright}
\centering
\setlength{\tabcolsep}{1em} 
\begin{tabular}{l c c c}
\toprule
	 Recipe & reflow &  no reflow &  previous \\\midrule
     Coincidences & 8800(300)     & 1490(60)  & 600(100)\\
\bottomrule
\end{tabular}
\caption{Comparison of coincidence counts measured per second per mm waveguide length at 1\,mW external pump power for the three different fabrication recipes. The value for the previous recipe is taken from Ref. \cite{Auchter2021}.}
\label{tab:PDCData}
\end{table}

\section{Conclusion and outlook}
We optimized the fabrication recipe for BRWs to yield samples with lower optical loss and therefore higher photon pair production rates.
By employing FBMS optical lithography, low pressure and low chlorine concentration etching, and resist reflow, we reduced the RMS area sidewall roughness in test GaAs sample waveguides from 16.05(2)\,nm to 4.736(7)\,nm.
For our BRWs, the improved etch recipe results in a low optical loss coefficient for telecom wavelength light of \( \alpha_\mathrm{reflow}=0.08\left(6\right)\,\mathrm{mm}^{-1}\).
The lower optical losses lead to higher coincidence rates between the signal and idler photons created in PDC.
The rate of \(8800\left(300\right)\,\left(\mathrm{mW\cdot s\cdot mm}\right)^{-1}\) is around a six-fold increase compared to samples produced without the reflow step and a factor of around 15 better than previous samples.
This shows that a fabrication recipe benefits from optimization of every manufacturing step.
Resist reflow is a particularly important step in reducing the sidewall roughness of waveguides and is one of the main factors contributing to optical losses in miniaturized optical devices. 
The improved fabrication and resulting low loss BRWs testify to the adequacy of the AlGaAs platform for quantum communication applications.

\begin{backmatter}
\bmsection{Funding}
The authors acknowledge funding by the Uniqorn project (Horizon 2020 grant agreement no. 820474) and the BeyondC project (FWF project no. F7114).

\bmsection{Acknowledgments}
We thank Felix Laimer and Elisabeth Gruber for help with atomic force microscopy and Markus Weiss for support and fruitful discussions in the cleanroom. 

\bmsection{Author contributions}
Conceptualization, H.T., R.J.C., S.F., G.W.; Formal analysis, H.T., M.W., S.F.; Methodology,  H.T., M.W., R.J.C., S.F.; Investigation, H.T., M.W., B.N., A.S.; Resources,  H.S., M.K., S.H., C.S.; Software, H.T., S.F.; Supervision, R.J.C., S.F., C.S., G.W.; Writing - original draft, H.T.; Writing - review \& editing, All Authors; Funding acquisition, C.S., G.W.

\bmsection{Disclosures}
The authors have nothing to disclose.

\bmsection{Data availability} Data underlying the results presented in this paper are available at 10.5281/zenodo.7702404.

\end{backmatter}

\bibliography{Bibliography_FabPaper}

\end{document}